# Design and Implementation of Fixtures for Milling, Shaping and Drilling Operations

**Abdullah Al Hossain Newaz[1,*], Refat Jahan[2]**

[1]Mechanical Engineering, University of Bridgeport, Bridgeport, CT, USA
[2]Mechanical Engineering, University of New Haven, CT, USA
*Corresponding author: anewaz@my.bridgeport.edu

**Abstract**  This paper explores the production of a specified object using a combination of machining processes, including milling, shaping, and drilling, while emphasizing the critical role of fixture design in ensuring precision, repeatability, and efficiency. The study outlines the systematic approach to transforming raw materials into a finished product through an optimized manufacturing sequence that minimizes waste and maximizes productivity. Different machining operations are carefully selected based on their suitability for achieving the required dimensional accuracy and surface finish. Fixtures play a crucial role in maintaining workpiece stability, reducing vibration, and ensuring accurate alignment during machining. This paper discusses the selection and design of fixtures, such as jigs and V-blocks, which enhance positioning accuracy and contribute to consistent machining outcomes. Additionally, the integration of CNC technology is examined, highlighting its advantages in automation, process control, and precision enhancement. The proposed methodology not only improves the efficiency of material removal but also ensures compliance with quality standards, reducing machining errors and minimizing rework. Furthermore, potential advancements in fixture automation using pneumatic actuators are considered to further streamline operations. The findings of this study provide valuable insights into optimizing machining sequences and fixture design to achieve a cost-effective and high-quality manufacturing process.

## 1. Introduction

Manufacturing is the process of converting raw materials into finished products using tools, human labor, machinery, and chemical processing [1]. It plays a vital role in industrial growth by adding value to raw materials, enhancing their usability, and increasing market value [2].

This process involves various machining operations such as turning, drilling, chamfering, milling, and shaping, each chosen based on the required precision and product specifications [3]. The selection of appropriate machining methods ensures efficiency, minimal material waste, and high-quality output [4].

Different machines are used for specific manufacturing tasks, including lathe machines for rotational machining, milling machines for contouring and shaping, and drilling machines for hole-making operations [5]. Each machine has unique capabilities, making machine selection a critical factor in optimizing production [6].

With advancements in CNC technology, modern manufacturing has become more precise and automated, reducing errors and improving efficiency [7]. This paper explores the machining processes, machine selection, and fixture design necessary to achieve accurate and efficient manufacturing of a given object.

## 2. Detail Machine Selection, Manufacturing Sequence & Fixture Design

### 2.1. Machine Selection

For the completion of this job, we will utilize various machining processes, each chosen for its ability to achieve the required precision and efficiency.

### 2.2. Milling Machine

The milling process removes material through multiple small, precise cuts. This is achieved by using a multi-toothed cutter that rotates at high speed while the workpiece is either advanced slowly or remains stationary, depending on the desired outcome. In most cases, a combination of high-speed rotation, controlled feed rate, and optimized cutting paths ensures accuracy and surface finish [2,3], Figure 1 [11].

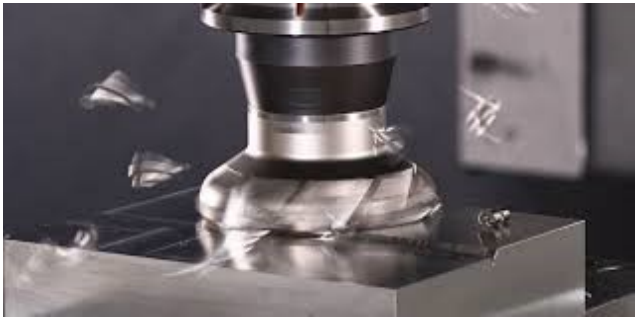

**Figure 1.** Milling Machine

## 2.3. Shaping Machine

The shaping process is a metal removal technique in which a single point cutting tool is mounted on a reciprocating ram. This tool moves in a linear direction across the workpiece, which is securely held on the machine table. Material is removed in horizontal, vertical, and angular planes as the tool cuts during the forward stroke, while the return stroke remains idle to enhance efficiency [3], Figure 2 [12]

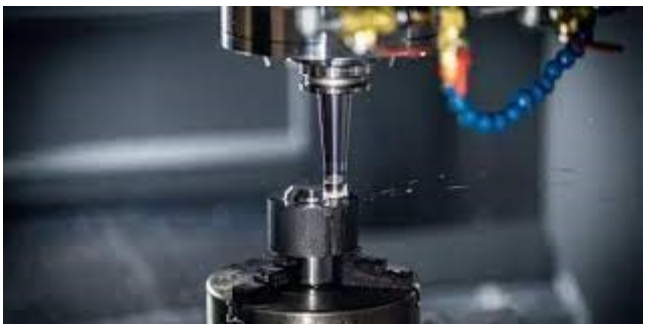

**Figure 2.** Shaping Machine

## 2.4. Drilling Machine

A drilling machine operates by rapidly rotating a cutting tool (drill bit) and progressively lowering it into the workpiece at a controlled speed and feed rate to create precise cylindrical holes [3]. To ensure accuracy and safety, the workpiece must be firmly secured on the drill table using vises and clamps, preventing movement during the drilling process. Figure 3 [13].

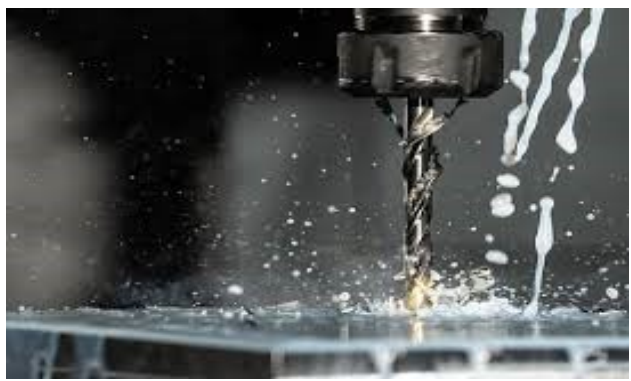

**Figure 3.** Drilling Machine

## 2.5. Manufacturing Sequence for the Given Objective

### 2.5.1. Object: 2 x R 2.5"

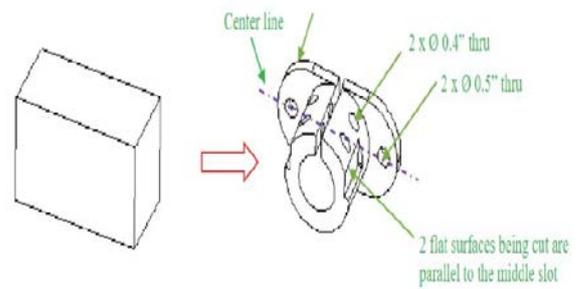

**Figure 4.** Object Hand drawing with dimension

### 2.5.2. Manufacturing Sequence

**Milling operation**

↓

**Contour Milling Operation**

↓

**Face Milling Operation**

↓

**Center Drilling Operation**

↓

**End Milling Operation**

↓

**Drilling Operation**

## 3. Fixture Design

Our object:

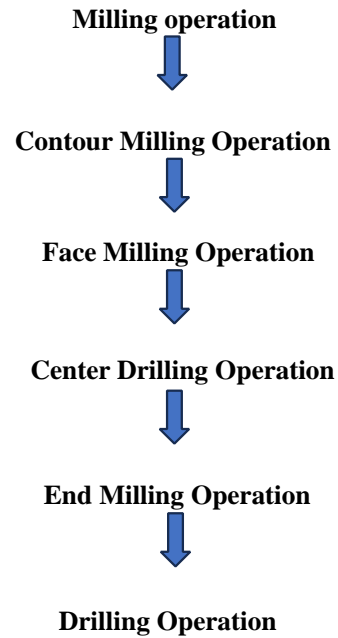

**Figure 5.** Object

This object is drawn by Creo 8.0.

*Begin material removal using a CNC milling machine, following the manufacturing sequence.

*Use the machine to shape the part accurately.

*The process will result in the initial shape of the component

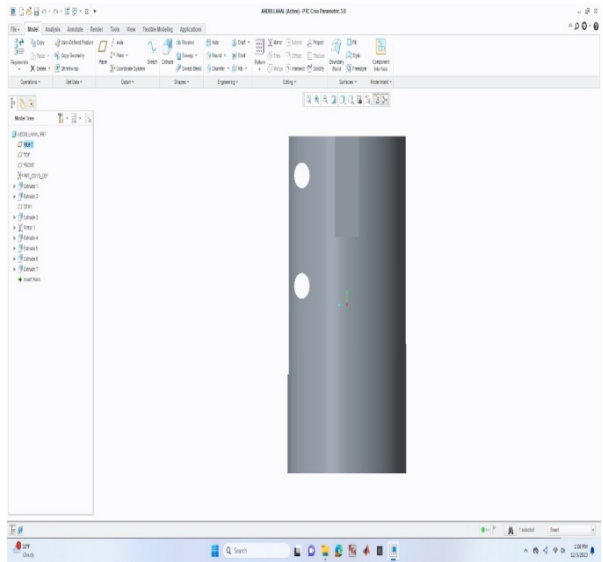

**Figure 6.** object two holes

*For the center drilling process, we will hold the object using a jig fixture and drill vertically to create holes through the cylinder.

*Next, we will create a slot at the center of the cylinder using V-Block fixtures, as shown in Figure 7.

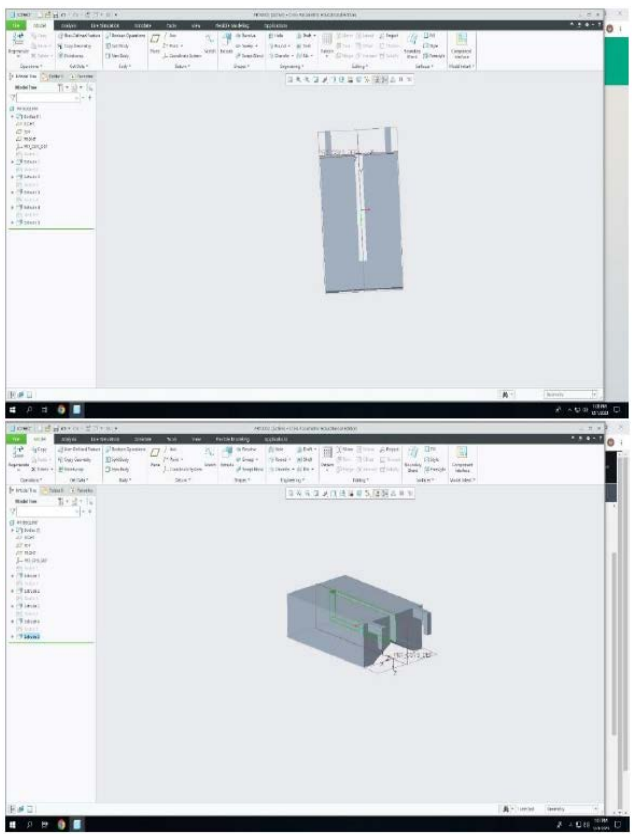

**Figure 7.** V-Top supporting for doing holes

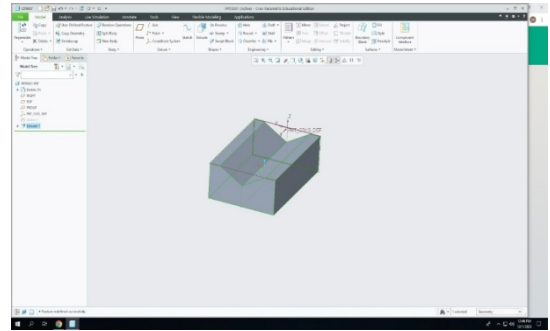

**Figure 8.** V-Bottom for doing holes

For the next steps, we will need to fix the object in a different fixture specifically designed for these operations. This fixture will ensure proper alignment and stability during the upcoming processes.

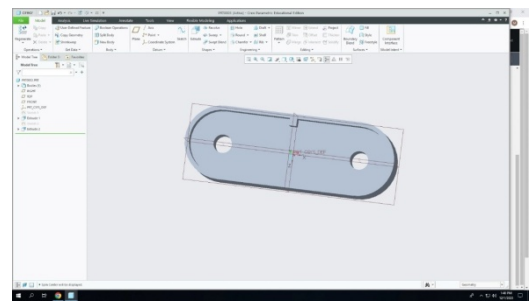

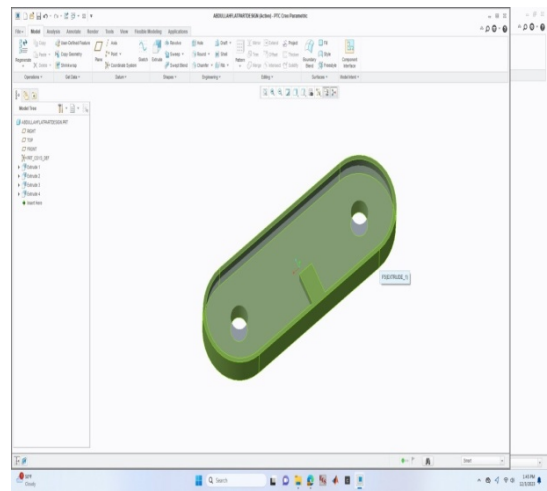

**Figure 9.** Fixture design to fix the object

Our next step is to create two flat surfaces that will be parallel to the middle slot. To achieve this, we can use the different V-block fixtures shown Figure 10.

To ensure precise positioning during the manufacturing process, we will utilize a fixture with key alignment. This will guarantee that the object is correctly positioned, with the slots parallel to the center slot. The next step involves using a milling machine to create two flat surfaces on the component.

For the following operation, we will continue using the same fixture to ensure the two holes are drilled perpendicular to the center slots. This will be achieved through vertical drilling. Additionally, horizontal drilling will be used to create two reference holes on the fixture. These reference holes will guide the drilling of two holes on the flat surface, ensuring they are accurately centered within the circular part of the flat surface.

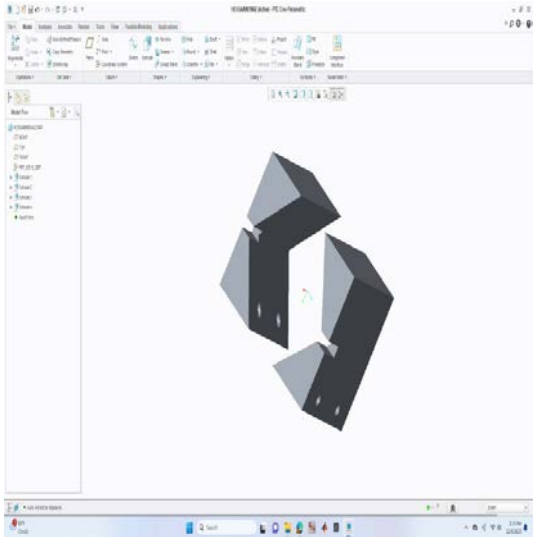

**Figure 10.** V-shaped fixture

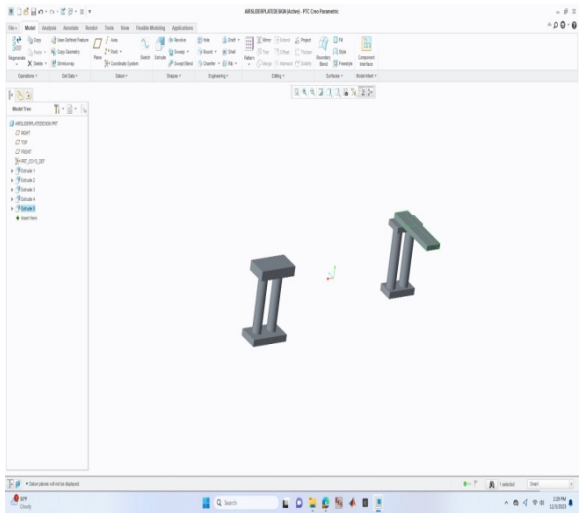

**Figure 11.** Air slider

An air slider and plate with a block are used to accurately position the cylinders through hole. The air slider also supports the V-block, ensuring stable alignment and precise positioning throughout the process

## 4. Possible Future Improvement: [5,7,8]

**Improved Fixture Layout Design:** Optimizing the fixture layout can significantly enhance the dimensional and form accuracy of the workpiece. By minimizing deformation during machining, the workpiece will be held more securely, which results in higher precision and reduces the risk of errors caused by misalignment or shifting. This will contribute to the overall quality of the final product.

**Flexible Fixtures:** In industries that deal with a wide range of products and frequently changing designs, flexible fixtures are becoming indispensable. Their ability to quickly adapt to different part geometries reduces setup times and ensures that production can proceed without delays. This flexibility also supports shorter production runs, ultimately leading to greater cost-efficiency and faster time-to-market.

**Automated Jigs or Fixtures:** By designing a fixture or jig that can automatically hold a part during secondary operations, the need for human intervention can be minimized. This not only improves the speed and consistency of the process but also enhances safety by reducing direct handling of components. It also ensures that the part is held securely and positioned correctly without the risk of human error.

**Pneumatic Actuation for Automation:** Automating the fixture using pneumatic actuators offers a significant advantage by enabling faster, more precise adjustments and reducing manual labor. Pneumatic actuators can precisely control the positioning of the workpiece, ensuring consistency across operations. This also eliminates variability caused by human error and improves operational efficiency, making the entire process more streamlined and reliable.

## 5. Conclusion

In conclusion, the manufacturing of the given job using a CNC milling machine has been successfully carried out, achieving a high level of precision and efficiency. Through the execution of various operations, including face milling, end milling, and drilling, we were able to shape the workpiece to the required specifications.

The face milling operation, involving rough, finish, and contour cuts, utilized different sizes of tools, such as rough mills, contour mills, and finish mills. This process ensured the desired surface quality and part geometry. Additionally, the drilling operation, performed using multiple types of drill mills, allowed for precise hole sizes, essential to the final design.

A key aspect of this project was the design and implementation of a custom fixture, specifically created to hold the workpiece securely during machining. The fixture was engineered with careful consideration to avoid interference with machining operations and tooling. Moreover, its flexible design allowed it to adapt seamlessly to the various operations performed, ensuring smooth execution and minimal downtime.

The uniqueness of this process lies in the tailored fixture design, which not only enhanced operational efficiency but also enabled greater adaptability across multiple machining stages. This approach demonstrates how precise fixture design, combined with advanced CNC machining, can deliver exceptional results, ensuring both quality and flexibility in manufacturing

## References

[1] "Modular Fixture Design for CNC Machining Centers", Yusuf Kılıçarslan.
[2] "Optimization of Machining Processes Preparation with Usage of Strategy Manager", M. Pivarčiová, M. Frankovský, M. Kenderová.
[3] "Design, Dimensioning, and Optimization of a Fixture for CNC Machining of a Housing Part",João Pedro Ferreira.
[4] "Design of Machining Fixture for Support Bracket", Josco Jose, Kiran Prakash, Harikrishnan K M, Sarath Chandran M, Midhun K M, Job Jose P.
[5] "Numerical Control Machine Optimization Technologies through Digital-Twin Technology",Kazuhiro Iwata, Takashi Matsumura, Hiroshi Yamamoto.

[6] "Research on Optimization of NC Machining Parameters", S. S. Mahapatra, Amar Patnaik.

[7] "A Case-Based Reasoning Approach for Design of Machining Fixture", S. S. Pande, S. S. Pande.

[8] "Design and Development of Fixture for CNC: Reviews, Practices, Future Directions", S. S. Pande, S. S. Pande.

[9] "Research and Application of CNC Machining Method Based on CAD/CAM/Robot Integration", Jianhua Zhang, Yujie Li, Xiaodong Zhang.

[10] "Design of Modular Fixture for CNC Machining Center", S. S. Pande, S. S. Pande.

[11] Modeling and Optimization Process in Milling, Bogdan A. Chirita, Catalin I. Pruncu, Jun Jiang, 27 April 2021.

[12] Editorial: Shaping the future of materials science through machine learning, Dezhen Xue, Turab Lookman, 18 December 2024.

[13] Drilling mechanism of temperature-sensitive non-directional UHMWPE fiber composites Chaolong Fu, Yuyuan Huang, Qingrui Jiang, Yujie Ma, Shanshan Hu, Genge Zhang, 08 January 2025.